%% file: ms.tex
\documentclass[aps,prl,reprint,superscriptaddress,twocolumn,amsmath,amssymb,floatfix]{revtex4-2}

\usepackage[T1]{fontenc}
\usepackage{newtxtext} % publication fonts
\usepackage[smallerops,varbb,varg,subscriptcorrection,timesmathacc,uprightGreek]{newtxmath} % publication fonts
\usepackage{eucal} % publication fonts (calligraphy math)
\usepackage{bm}
\renewcommand\vec{\bm}

\usepackage{graphicx}
\usepackage[dvipsnames,svgnames,x11names,hyperref]{xcolor}
\usepackage{hyperref}
\hypersetup{colorlinks=true,linkcolor=NavyBlue,urlcolor=NavyBlue,  citecolor=NavyBlue}  % closer in appearance to what APS journals use

\usepackage{physics,braket,slashed}

\newcommand{\pdagger}{{\phantom{\dagger}}}
\usepackage{enumitem}

\newcommand{\orcid}[1]{\href{https://orcid.org/#1}{\includegraphics[width=8pt]{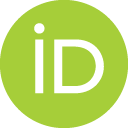}}}

\newcommand{\vop}{v}
\newcommand{\JJ}{J}
\newcommand{\vvop}{\vec{\vop}}
\newcommand{\sss}{\scriptscriptstyle}
\newcommand{\labelR}{\textrm{R}}
\newcommand{\labelJ}{\textrm{J}}
\newcommand{\labelK}{\sss K}
\newcommand{\labelKp}{\sss K'}
\newcommand{\NN}[1]{\langle #1 \rangle}
\newcommand{\NNN}[1]{\langle\langle #1 \rangle\rangle}
\newcommand{\indexI}{{\sss I}}
\newcommand{\indexJ}{{\sss J}}
\newcommand{\indexIJ}{\indexI \indexJ}
\newcommand{\nuIJ}{\nu_{\indexI\indexJ}}
\newcommand{\nuJI}{\nu_{\indexJ\indexI}}
\newcommand{\psiI}{\psi_{\indexI}}
\newcommand{\psiJ}{\psi_{\indexJ}}
\newcommand{\rhoI}{\rho^{\indexI}}

\newcommand{\lambdaR}{\lambda_{\sss\textrm{R}}}
\newcommand{\lambdaRi}{\lambda_{{\sss\textrm{R}},i}}

\newcommand{\val}{v}

\begin{document}
%\vspace*{0.8cm}

\input{main.tex}

%%%%%%%%%% Merge with supplemental materials %%%%%%%%%%
\onecolumngrid
\clearpage
\begin{center}
\textbf{\large Supplementary Material for \\ ``Spontaneous Valley Spirals in Magnetically Encapsulated Twisted Bilayer Graphene''}
\end{center}
%%%%%%%%%% Merge with supplemental materials %%%%%%%%%%
%%%%%%%%%% Prefix a "S" to all equations, figures, tables and reset the counter %%%%%%%%%%
\setcounter{equation}{0}
\setcounter{figure}{0}
\setcounter{table}{0}
\setcounter{page}{1}
\makeatletter
\renewcommand{\theequation}{S\arabic{equation}}
\renewcommand{\thefigure}{S\arabic{figure}}
\renewcommand{\bibnumfmt}[1]{[S#1]}
\renewcommand{\citenumfont}[1]{S#1}
%%%%%%%%%% Prefix a "S" to all equations, figures, tables and reset the counter %%%%%%%%%%

\input{main_supmat.tex}

\input{supmat.bbl}
\end{document}

%% file: main.tex
%!TeX root = ms.tex

\title{Spontaneous Valley Spirals in Magnetically Encapsulated Twisted Bilayer Graphene}

\author{Tobias M. R. Wolf \orcid{0000-0003-4665-9874}}
\author{Oded Zilberberg \orcid{0000-0002-1759-4920}}
\author{Gianni Blatter \orcid{0000-0003-0521-8028}}
\affiliation{Institute for Theoretical Physics, ETH Zurich, 8093 Zurich, Switzerland}

\author{Jose L. Lado \orcid{0000-0002-9916-1589}}
\affiliation{Department of Applied Physics, Aalto University, 00076 Aalto, Espoo, Finland}

\begin{abstract}
Van der Waals heterostructures provide a rich platform for emergent physics due
to their tunable hybridization of electronic orbital- and spin-degrees of
freedom. Here, we show that a heterostructure formed by twisted bilayer
graphene sandwiched between ferromagnetic insulators develops flat bands
stemming from the interplay between twist, exchange proximity and spin-orbit
coupling. We demonstrate that in this flat-band regime, the spin degree of
freedom is hybridized, giving rise to an effective triangular superlattice with
valley as a degenerate pseudospin degree of freedom. Incorporating electronic
interactions at half-filling leads to a spontaneous valley-mixed state, i.e., a
correlated state in the valley sector with geometric frustration of the valley
spinor. We show that an electric interlayer bias generates an artificial
valley--orbit coupling in the effective model, controlling both the valley
anisotropy and the microscopic details of the correlated state, with both
phenomena understood in terms of a valley-Heisenberg model with easy-plane
anisotropic exchange. Our results put forward twisted graphene encapsulated
between magnetic van der Waals heterostructures as platforms to explore purely
valley-correlated states in graphene.
\end{abstract}

\date{\today}

\maketitle
%%%%%%%%%%%%%%%%%%%%%%%%%%%%%%%%%%%%%%%%%%%%%%%%%%%%%%%%%%%%%%%%%%%%%
%%%%%%%%%%%%%%%%%%%%%%%%%%%%%%%%%%%%%%%%%%%%%%%%%%%%%%%%%%%%%%%%%%%%%
%%%%%%%%%%%%%%%%%%%%%%%%%%%%%%%%%%%%%%%%%%%%%%%%%%%%%%%%%%%%%%%%%%%%%

% % % % % % % % % % % % % % % % % % % % % % % % % % % % % % % % % %
% % % INTRODUCTION

Twisted graphene multilayers have risen as a paradigmatic platform
for engineering correlated states of matter. Their unique flexibility
stems from the emergence of a tunable length scale, the moir\'e
length, which generates electronic spectral minibands with a
controllable ratio between the kinetic and interaction energies. As a
result, a variety of strongly-correlated states appear in these
twisted van der Waals materials, such as intrinsic
superconductivity~\cite{CaoJarilloHerrero2018N, YankowitzDean2019S,
LuEfetov2019N}, strange metal
behavior~\cite{CaoJarilloHerrero2020PRL}, and correlated
insulators~\cite{CaoJarilloHerrero2018Na}. Furthermore, this platform
can realize correlated states that are rarely found in nature, such
as ferromagnetic superconductivity~\cite{CaoJarilloHerrero2020N} and
interaction-driven quantum anomalous Hall
effect~\cite{SerlinYoung2019S}.

The correlated states in twisted graphene multilayers that were explored thus
far mostly focus on spontaneous symmetry-breaking of the spin ($\pm 1/2$)
degree of freedom, i.e., of the symmetry group $SU(2)_s$
\cite{SanchezYamagishiJarilloHerrero2016NN}. Interestingly, low-energy charge
carriers in graphene also have two valleys ($K$, $K'$) as a well-defined
(spinor) quantum number with (approximate $SU(2)_\val$ \cite{ChoiAguado2005,
JarilloDeFranceschi2005nature, XuBalents2018PRL, NatoriAndrade2019PRB} or)
$U(1)_\val$ symmetry, which offers additional possibilities for spontaneous
symmetry breaking due to interactions, e.g., spontaneous valley-polarized
ground states~\cite{SerlinYoung2019S}. So far, however, interaction-induced
valley spatial textures have not been considered. Here, we show that
proximity-induced spin--orbit coupling can lock spin- and orbital degrees of
freedom in a way that generates exotic symmetry breaking in the valley sector
when electronic interactions are included.

%%%%%%%%%%%%%%%%%%%%%%%%%%%%%%%%%%%%%%%%%%%%%%%%%%%%%%%%%%%%%%%%%%%%%
\begin{figure}%[htb]%[!t]
\includegraphics[width=3.41in]{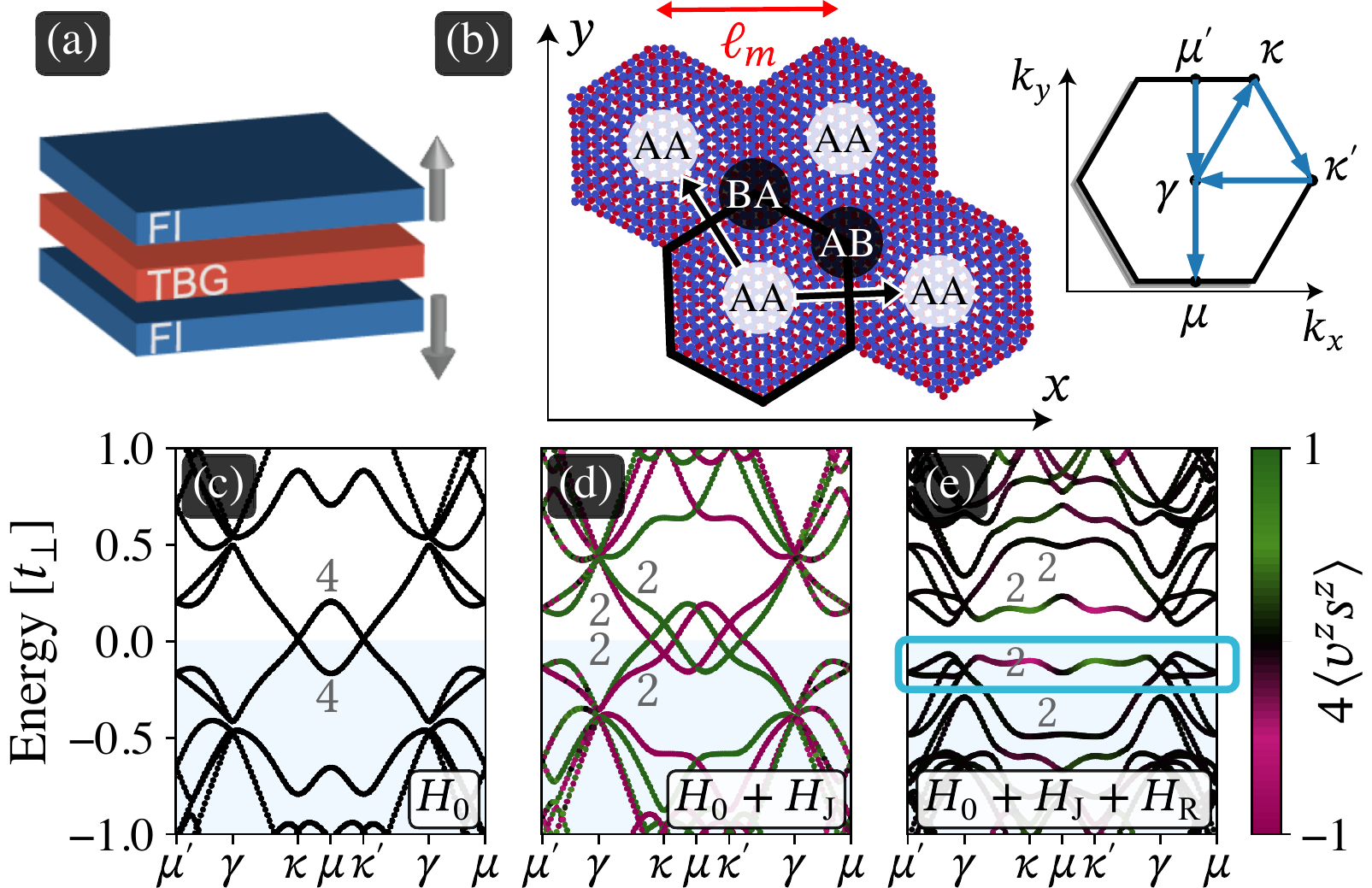} %3.41in
\caption{\label{fig:fig1}
Structure and single-particle electronic properties of twisted bilayer graphene
(TBG) encapsulated within ferromagnetic insulators (FI).  (a)~Sketch of the
encapsulated system, where arrows denote the magnetization orientation of each
FI. (b)~Moir\'e spatial pattern arising from stacking two graphene layers with
relative twist angle $\alpha$. The pattern has a length scale $\ell_m$ with
characteristic AA, AB, and BA regions. It generates a hexagonal mini-Brillouin
zone with characteristic high-symmetry points. (c-e) Bandstructures at twist
angle $\alpha\simeq 2^{\circ}$, interlayer coupling $t_\perp = 0.12t$, and no
interlayer bias ($V=0$) along the high-symmetry path
$\gamma$--$\kappa$--$\kappa'$--$\gamma$--$\mu$: for the isolated TBG~(c),
including local exchange fields with $m=t_\perp/3$~(d), and including both
local exchange fields with $m=t_\perp/3$ and Rashba SOC
$\lambdaR=t_\perp/3$~(e). The coloring of the bands indicates the expectation
value of the valley--spin operator $\langle \vop^z s^z \rangle$, showing
fixed spin and valley in (d) and finite spin-mixing at fixed valley in (e).
The light-blue box marks the flat band below charge neutrality.
}
\end{figure}
%%%%%%%%%%%%%%%%%%%%%%%%%%%%%%%%%%%%%%%%%%%%%%%%%%%%%%%%%%%%%%%%%%%%%

Spin--orbit coupling effects in monolayer graphene lead to the
emergence of the quantum anomalous Hall effect~\cite{QiaoNiu2010PRB,
WangShi2015PRL}. They are tuned experimentally  using electric
fields~\cite{GuimaraesWees2014PRL} and by proximity to
semiconductors~\cite{SafeerCasanova2019NL}. Note that Rashba SOC
effects can be on the order of $0.1$~meV in single-layer graphene
encapsulated in Boron-Nitride~\cite{GuimaraesWees2014PRL}, $0.3$~meV
in hydrogenated graphene~\cite{GmitraFabian2013PRL}, and up to
$1.5$~meV for graphene on dichalcogenides~\cite{SafeerCasanova2019NL,
YangShi2017PRB}. Moving to  twisted graphene bilayers, this energy
scale should be compared with a typical Coulomb correlation gap of
$\sim 0.3$~meV~\cite{CaoJarilloHerrero2018Na}. Remarkably, even
though the Rashba SOC can compete with these correlated gaps, this
interplay has thus far not received much attention in twisted van der
Waals materials.

In this work, we focus on the valley degree of freedom, described as a
two-spinor, and demonstrate the emergence of correlations in the valley spinor
of twisted bilayer graphene encapsulated within ferromagnetic insulators (FIs),
such as $\mathrm{CrI}_3$, see Fig.~\ref{fig:fig1}(a). We show that the
combination of twist engineering alongside proximity-induced magnetic exchange
and Rashba spin--orbit coupling hybridizes the spin degree of freedom and leads
to valley-degenerate flat bands. It is this valley-degeneracy in the absence of
spin-degeneracy that provides us with a unique playground for symmetry-broken
states solely in the valley sector.  To describe the latter, we propose a
phenomenological triangular lattice model that captures the low-energy
flat-band valley-physics.  At half-filling of the flat bands, we find that
screened Coulomb interactions lead to a symmetry breaking with valley-spiral
order. Furthermore, we find that the latter is described by an anisotropic
valley-Heisenberg model and that the easy-axis anisotropic valley-exchange can
be controlled through electric interlayer bias. Finally, we discuss potential
experimental scenarios to detect this effect.

% % % % % % % % % % % % % % % % % % % % % % % % % % % % % % % % % %
% % % % % % % % Main text

Our system consists of twisted bilayer graphene encapsulated in the
$z$-direction between ferromagnetic insulators, see Fig.~\ref{fig:fig1}(a). We
describe the electronic properties of the system using an effective atomistic
tight-binding Hamiltonian for the graphene bilayer
\begin{align} \label{eq:fullModel}
H=H_0+H_{\labelJ}+H_{\labelR},
\end{align}
where the electronic degrees of freedom of the FI are integrated out.  The
Hamiltonian $H_0$ describes the bare twisted bilayer, $H_{\labelJ}$ includes
proximity-induced exchange fields (induced by virtual tunneling processes
between the bilayer and the FI)~\cite{ZhongXu2017SA, YangChshiev2013PRL,
SinghKawakami2017PRL, ZollnerFabian2019PRB, PeraltaMireles2019PRB,
HanFabian2014NN, YangChshiev2013PRL,WangShi2015PRL}, and $H_{\labelR}$
contributes a Rashba spin--orbit coupling that stems from a combination of
proximity-induced spin--orbit coupling and locally-broken mirror
symmetry~\cite{SafeerCasanova2019NL, YangShi2017PRB, DedkovLaubschat2008PRL}.
The bare Hamiltonian of the bilayer reads
$
 H_{0} = \sum_{\NN{i,j},s} t\, c^{\dagger}_{i,s} c^{\pdagger}_{j,s}
          +  \sum_{i,j,s} t^{\perp}_{ij} \, c^{\dagger}_{i,s}c^{\pdagger}_{j,s}
          - \sum_{i,s}V_i\,c^{\dagger}_{i,s} c^{\pdagger}_{i,s} ,
$
where $c_{i,s}^{(\dagger)}$ destroys (creates) an electron with spin
${s\in\lbrace\pm 1/2\rbrace}$ at position ${\vec{r}_i=(x_i,y_i,z_i)}$ in one of
the layers located at ${z_i=\pm d/2}$. We consider intralayer nearest-neighbor
hopping with amplitude ${t\simeq 2.7\,\text{eV}}$~\cite{CastroNetoGeim2009RMP}.
The interlayer hopping from site $\vec{r}_i$ to $\vec{r}_j$ is parametrized as
${t^{\perp}_{ij} = t_{\perp} [(z_{i}-z_{j})^{2}/|\vec{r}_{i} -
\vec{r}_{j}|^{2}] \, e^{-(|\vec{r}_{i}-\vec{r}_{j}|-d)/\ell} }$  with
${t_{\perp}\simeq 0.12\,t}$ that describes the hybridization over the
interlayer distance ${d\simeq 2.35 a_0}$ with $a_0$ the intralayer bond length
and $\ell\simeq 0.3\,a_0$ controlling the interlayer hopping range
\cite{ReichOrdejon2002PRB, LaissardiereMagaud2010nano, MoonKoshino2013PRB}. The
onsite potentials $V_i=\mu+\mathrm{sgn}(z_i) V$ describe the overall chemical
potential $\mu$ and electric interlayer bias $V$.

We first discuss the system in the absence of interlayer bias, $V=0$. Each
isolated graphene layer $l\in\{1,2\}$ exhibits a characteristic spectrum with
Dirac-like band touchings at valleys $K, K'$~\cite{CastroNetoGeim2009RMP},
which we label with the eigenvalues $v\in\{\pm 1/2\}$ of the valley operator
$v^z$, respectively~\cite{ColomesFranz2018PRL, RamiresLado2018PRL,
RamiresLado2019PRB,WolfZilberberg2019PRL}. Consequently, the decoupled bilayer
has spectral bands that are eightfold degenerate, characterized by layer,
valley, and spin indices, $\ket{l, \val, s}$, respectively. Interlayer coupling
($t_{\perp}\neq0$), mixes the energy bands between the layers. Furthermore, a
twist angle $\alpha$ between the layers leads to a moir\'e superlattice
structure with a characteristic distance $\ell_m$ and regions labeled AA- and
AB/BA in accord with the alignment of the A and B sites of each graphene layer
on top of each other, see Fig.~\ref{fig:fig1}(b). The resulting large
superstructure implies that the electronic spectrum of $H_0$ consists of many
minibands, resulting from backfolding the dispersion of each graphene layer and
subsequent hybridization by the interlayer
coupling~\cite{SboychakovNori2015PRB}, see Fig.~\ref{fig:fig1}(c).  For a large
moir\'e length $\ell_m$ and low energies, intervalley scattering can be
neglected, i.e., $t_{\perp}$ does not couple different valleys. As a result,
each miniband at Bloch momentum $\vec{k}$ (corresponding to valley $K$) is
degenerate in spin and has a valley-partner at $-\vec{k}$ (corresponding to
valley $K'$). Hence, each eigenvalue is at least four-fold
degenerate~\cite{LopesdosSantosCastroNeto2012PRB,
SuarezMorellBarticevic2010PRB, LopesdosSantosCastroNeto2007PRL,
BistritzerMacDonald2011PotNAoS,SanJosePrada2013PRB}, or higher along
high-symmetry lines in the mini-Brillouin zone (mBZ). Crucially, except for
fine-tuned angles~\cite{SuarezMorellBarticevic2010PRB,
BistritzerMacDonald2011PotNAoS,WolfZilberberg2019PRL, KoshinoFu2018PRX,
KangVafek2018PRX, PoSenthil2018PRX,HejaziBalents2019PRB} or in the limit of
tiny twist angles~\cite{SanJosePrada2013PRB, RamiresLado2018PRL}, the
low-energy minibands are typically dispersive.

The encapsulation of the TBG between ferromagnetic insulators with
magnetization pointing out of plane (and an antiferromagnetic alignment
between the FIs) [cf.~Fig.~\ref{fig:fig1}(a)] profoundly alters the low-energy
spectrum. In this configuration, the FIs induce exchange fields with effective
moment $\vec{m}_i=\mathrm{sgn}(z_i) m \,\vec{\hat{z}}$ at each site~$\vec
r_i$, and the locally-broken mirror symmetry generates Rashba spin--orbit
coupling $\lambdaRi=\mathrm{sgn}(z_i)\lambdaR$ in each graphene
layer~\cite{GongZhang2019S}. These effects are, respectively, described by
$
\smash{ % makes sure the formula does not affect line spacing
H_{\labelJ} = \sum_{j, s s'} (\vec{m}_{j}
  \cdot \vec{\sigma})_{s s'}  \; c^{\dagger}_{j, s} c^\pdagger_{j,s'},
}
$
and
$
\smash{ % makes sure the formula does not affect line spacing
H_{\labelR} = i  \sum_{\NN{i,j},s s'} \lambdaRi
\,(\vec{\sigma} \times \vec{d}_{i j})^z_{ss'} \, c^{\dagger}_{i,s}
c^\pdagger_{j,s'} ,
}
$
where $\vec{d}_{i j}$ is the bond vector connecting intralayer sites $i,j$,
and the components of $\vec{\sigma} = (\sigma^x, \sigma^y, \sigma^z)$ are the
Pauli matrices describing spin.  As we assume the two FI layers to be
antiferromagnetically-aligned along the $z$-axis, the induced exchange fields
act as a (spin-dependent) magnetic interlayer
bias~\cite{CardosoFernandezRossier2018PRL}. Interestingly, even though the
exchange field $\vec{m}_i$ breaks time-reversal symmetry, the eigenstates
remain spin degenerate, see Fig.~\ref{fig:fig1}(d).  This is a result of the
symmetric orbital distribution between the layers, i.e., the spin-$\uparrow$
bands of one layer remain degenerate with the spin-$\downarrow$ bands of the
other (and vice versa), while the interlayer coupling does not mix spins. The
Rashba coupling term $\lambdaR$, however, mixes the two spin channels,
introduces a sizeable hybridization gap around charge neutrality, and
flattens-out the otherwise dispersive bands, see Fig.~\ref{fig:fig1}(e).

\begin{figure}%[t!]
\includegraphics[width=\columnwidth]{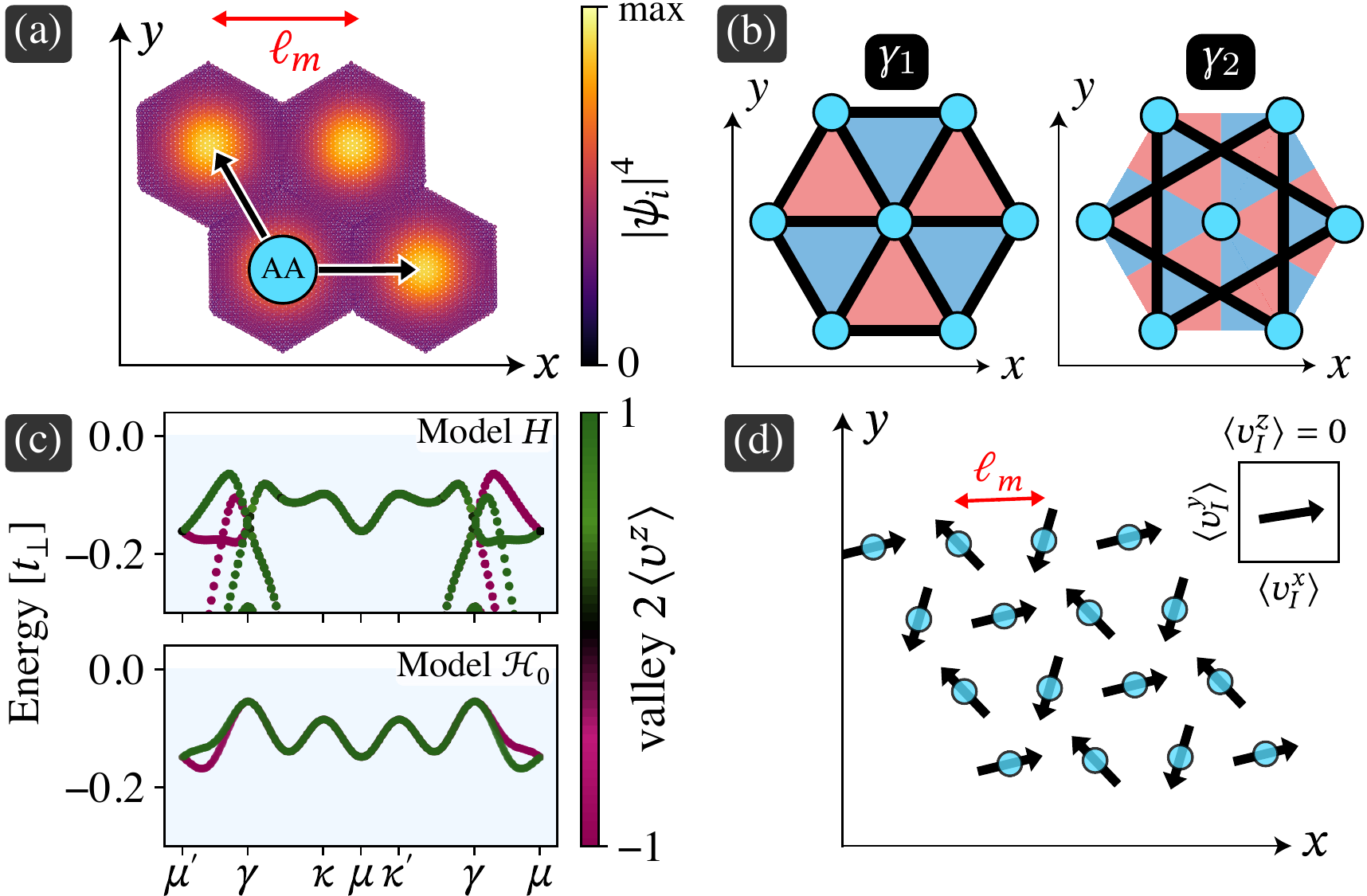}
\caption{\label{fig:fig2}
Effective triangular lattice model for the moir\'e orbitals of the flat band.
(a) Local density of states of the flat band below charge neutrality
highlighted in Fig.~\ref{fig:fig1}(e).
(b) A sketch of the triangular lattice model $\mathcal{H}_0$, see
Eq.~\eqref{eq:effmodel}, where cyan circles represent the AA regions, the
black lines denote first- and second-neighbors hoppings $\gamma_{1}$,
$\gamma_2$ respectively, and the red/blue triangles represent the staggered
flux patterns associated with the phases $\phi_1$ and $\phi_2$ for first- and
second-neighbor hopping.
(c, top)~Close-up on the flat band in Fig.~\ref{fig:fig1}(e), and (c, bottom)
comparison with the band of the phenomenological model [cf.
Eq.~\eqref{eq:effmodel} with $\gamma_1/t_\perp= 0.03$, $\gamma_2/t_\perp=0.09$,
$\phi_1=0$, $\phi_2=-0.4$]. The band color indicates the valley index
$v=\langle\vop^z\rangle$ and illustrates the valley-degeneracy along
$\gamma$--$\kappa$--$\kappa'$--$\gamma$ (green on top of magenta).
(d)~(In-plane) valley spiral appearing in the mean field ground state of
$\mathcal{H}_0+\mathcal{H}_U$, cf.~Eqs.~\eqref{eq:effmodel} and
\eqref{eq:effmodelInt}.  Arrows illustrate the valley polarization
$\langle\vvop_I\rangle$ of the respective orbital (inset).
}
\end{figure}

As a result, the FI-encapsulated twisted bilayer features a
pronounced van Hove singularity adjacent to the energy gap at charge
neutrality. This singularity becomes most pronounced for a fine-tuned
value of the ratio $\alpha/t_\perp$ between twist angle and the
interlayer coupling, here corresponding to physical parameters
$\alpha\approx 2^{\circ}$,  $t_\perp=0.12t$, % \alpha=\alpha_{16,1}
when $m=t_\perp/3$, $\lambdaR=t_\perp/3$~\cite{supmat}. The corresponding bands
then become maximally flat, see Fig.~\ref{fig:fig1}(e), and their wavefunctions
are mostly concentrated within the AA region of the moir\'e unit cell, see
Fig.~\ref{fig:fig2}(a). Importantly, these bands are only two-fold degenerate
in the \textit{valley} degree of freedom, whereas spin degeneracy is fully
broken -- in contrast with other graphene multilayer systems, where spin- and
layer-degeneracies persist~\cite{LopesdosSantosCastroNeto2007PRL}.

%%%%%%%%%%%%%%%%%%%%%%%%%%%%%%%%%%%%%%%%%%%%%%%%%%%%%%%%%%%%%%%%%%%%%
%%%%%%%%%%%%%%%%%%%%%%%%%%%%%%%%%%%%%%%%%%%%%%%%%%%%%%%%%%%%%%%%%%%%%

Crucial to our work, these low-energy flat bands resemble a simple effective
model for hopping between Wannier moir\'e orbitals arranged in a triangular
superlattice, see Fig.~\ref{fig:fig2}(b), i.e.,
\begin{align} \label{eq:effmodel}
\mathcal{H}_0 = \! \sum_{\sss\NN{\indexIJ}} \gamma_1\,\psiI^\dagger e^{i
\sigma^z \nuIJ \phi_1}  \psiJ + \! \sum_{\sss\NNN{\indexIJ}} \gamma_2\,\psiI^\dagger
 e^{i \sigma^z \nuIJ \phi_2}  \psiJ ,
\end{align}
with the valley spinors $\psiI^{(\dagger)} = (d_{\indexI,1/2}^{(\dagger)},
d_{\indexI,-1/2}^{(\dagger)})$ and destruction (creation) operators
$\smash{d_{\indexI,\val}^{(\dagger)}}$ for electrons on moir\'e unit cells $I$
with valley index $v$ taking the role of a pseudospin.  The form of the hopping
amplitudes follows from symmetry arguments \cite{supmat}, and we include first-
and second-neighbor amplitudes $\gamma_{1,2}>0$ with phases $\phi_{1,2}$, and
signs ${\nuIJ=-\nuJI\in\lbrace\pm 1\rbrace}$ that ensure symmetry under
rotation by $2\pi/3$, see Fig.~\ref{fig:fig2}(b).  Similar complex-valued
hopping amplitudes appear in the Kane-Mele model~\cite{KaneMele2005PRL} due to
spin--orbit coupling, such that we refer to $\phi_{1,2}$ as `valley--orbit
phases' in our model by analogy.  In the absence of interlayer bias, symmetry
enforces real first-neighbor hopping ($\phi_1=0$) \cite{supmat}, whereas
$\phi_2$ is finite in general. The hopping parameters can then be chosen to
qualitatively reproduce the flat band, see Fig.~\ref{fig:fig2}(c). We will see
how interlayer bias affects this low-energy valley-spinor model later.

The presence of a van Hove singularity (flat bands) naturally raises
the question how interactions affect the corresponding
electronic states near half-filling of the flat band. In the bilayer,
this corresponds to doping the system with one electron/hole per
moir\'e unit cell. Coulomb interactions in the microscopic model
\eqref{eq:fullModel} lead to effective Coulomb interactions
between the moir\'e orbitals in the low-energy model
\eqref{eq:effmodel}. Assuming that the screened Coulomb interaction
between the atoms is shorter-ranged than the moir\'e length scale
$\ell_m$ \cite{GuineaNiels2018PNAS}, the effective interaction
between the moir\'e orbitals becomes
\begin{align} \label{eq:effmodelInt}
\mathcal{H}_U = \frac{U}{2} \sum_{\indexI,\val} n_{\indexI,\val} n_{\indexI,-\val},
\end{align}
where $n_{\indexI,\val}=d^\dagger_{\indexI,\val}d^{\pdagger}_{\indexI,\val}$ is
the number operator for valley $\val$ of the moir\'e orbital $I$ and $U\simeq
0.15 t_\perp$ \cite{GuineaNiels2018PNAS} is the Hubbard interaction strength.
Our effective model $\mathcal{H} =
\mathcal{H}_0 + \mathcal{H}_U$ differs from the conventional Fermi-Hubbard
model~\cite{HubbardPRSL} in two respects: First, we have valley as pseudospin,
and second, our hopping amplitudes are complex. In what follows, we consider
half-filling such that the expected occupation number is ${\langle n_I \rangle
=1}$, and calculate the valley order of the ground state of $\mathcal{H}$.
Analogous to spin order, we characterize valley order by the expectation value
of the valley operator $\vec{\vop}_{\indexI} = \psiI^\dagger \vec\sigma
\psiI/2$ in each moir\'e cell $I$. We can interpret $\langle \vop_I^z \rangle$
as the local valley imbalance and $\langle \vop_I^{x,y} \rangle$ as local
valley coherence.
We will see that, similar to other spin-$1/2$ triangular lattice
models~\cite{SahebsaraSenechal2008PRL, VaeziHosseini2012PRB,
LiLi2016PRB,MisumiOhta2017PRB}, our model $\mathcal{H}$, cf.~Eqs.~\eqref{eq:effmodel} and~\eqref{eq:effmodelInt}, is prone to
valley-spiral states~[see Fig.~\ref{fig:fig2}(d)], and that valley--orbit
coupling, i.e., our complex-valued hoppings, can promote anisotropic
exchange~\cite{VaeziHosseini2012PRB}.

We determine the ground state using a self-consistent mean-field 
approximation for the many-body interaction, 
$
{\mathcal{H}_U\approx
\!  \sum_{\indexI} \psiI^{\dagger} \, \bar{U}(\rhoI) \; \psiI - E_0(\rhoI)}
$, 
where we introduced the density matrix
$
\rhoI = (\langle n_I
\rangle + \langle \vvop_I \rangle \cdot \vec{\sigma})/2
$ 
and the mean-field interaction ${\bar{U}(\rhoI)}$ and shift
${E_0(\rhoI)}$~\cite{supmat}.  Performing self-consistent relaxation of
different initial states, we find that interactions and geometrical frustration
in the triangular lattice favor a valley-spiral state on the length scale of
the moir\'e structure, see Fig.~\ref{fig:fig2}(d). We find that (i) the length
scale of the spiral varies slightly with the ratio $\gamma_2/\gamma_1$, and
(ii) that the spiral favors planar configurations with $\langle v_I^z
\rangle=0$.  Hence, the valleys seek a state with equal occupation ${\langle
n_{\indexI,\labelK} \rangle=\langle n_{\indexI,\labelKp} \rangle}$ and mix
coherently, $\langle
\vop_{\indexI}^{x,y} \rangle\neq 0$.  Interestingly, in the limit $\phi_2 \to
0$, stabilization of the in-plane spiral state is lost such that spiral states
with finite out-of-plane components $\langle v_I^z \rangle>0$ become degenerate
with in-plane spiral configurations; this suggests that the phases $\phi_1$ and
$\phi_2$ in $\mathcal{H}_0$ [see Eq.~\eqref{eq:effmodel} and Fig.\
\ref{fig:fig2}(b)] play a crucial role in defining the valley order.

To better understand our mean-field results, we expand the Hamiltonian
$\mathcal{H}$ at half-filling in the strong-interaction limit $U\gg
\gamma_1,\gamma_2$ using a Schrieffer-Wolff
transformation~\cite{SchriefferWolff1966PR,supmat} that takes us to a
valley-Heisenberg model with anisotropic and (anti-)symmetric exchange, i.e.,
\begin{align}\label{eq:Heisenberg}
   \mathcal{H}_{\vop} = \sum_{\indexIJ} \JJ_{\indexIJ} \, \vvop_{\indexI} 
   \!\cdot\! \vvop_{\indexJ} + \Delta_{\indexIJ} \, \vop_{\indexI}^z 
   \vop_{\indexJ}^z+ \nuIJ \, D_{\indexIJ}\,(\vvop_{\indexI} \times \vvop_{\indexJ})_z.
\end{align}
Here, $\JJ_{\indexIJ}$, $\Delta_{\indexIJ}$, and $D_{\indexIJ}$ denote the
isotropic, anisotropic, and antisymmetric exchange couplings, respectively.
These couplings are finite for first- and second-neighbor exchange only
(indexed by ${n = 1,2}$) and take the form $\JJ_n = \JJ_{n}^0\,
(\cos^2\phi_n-\sin^2\phi_n)$, $\Delta_{n} = 2 \JJ_{n}^0 \sin^2\phi_n$, and
$D_{n}=\JJ_{n}^0 \sin(2\phi_n)$, with $\JJ_{n}^0=2\gamma_{n}^2/U$. In the
absence of an interlayer bias (${V = 0}$), we have ${\phi_1 = 0}$ such that the
first-neighbor terms in $\mathcal{H}_{\vop}$ are isotropic. Generally, the
isotropic exchange couplings $J_n$ can turn valleymagnetic~\cite{note1}
(${J_n<0}$) as $\phi_n$ increases; however, for the regimes we consider here,
we can restrict ourselves to anti-valleymagnetic couplings (${J_n>0}$ for
${n=1,2}$), which favors valley spirals due geometric frustration in the
triangular lattice. The finite phase $\phi_2$ in the second-neighbor coupling
stabilizes in-plane valley configurations by inducing anisotropy ${\Delta_2>0}$
and favors second-neighbor valley misalignment (canting) due to the
antisymmetric coupling ${D_2>0}$. Note that the alternating nature of the signs
${\nuIJ\in\lbrace \pm 1\rbrace}$ in our triangular lattice favors valley
spirals as well, rather than chiral structures such as
skyrmions~\cite{HeinzeBluegel2011nature}. Consequently, there are two distinct
mechanisms driving valley spirals, such that the length scale of the valley
spiral depends on the competition between anti-valleymagnetic geometric
frustration ($J_n$, $n=1,2$) and the antisymmetric couplings ($D_n$, $n=1,2$).
In the following, we investigate how the addition of a finite electric
interlayer bias modifies the results discussed thus far.

%%%%%%%%%%%%%%%%%%%%%%%%%%%%%%%%%%%%%%%%%%%%%%%%%%%%%%%%%%%%%%%%%%%%%
\begin{figure}%[t!]%[!h]
\includegraphics[width=\columnwidth]{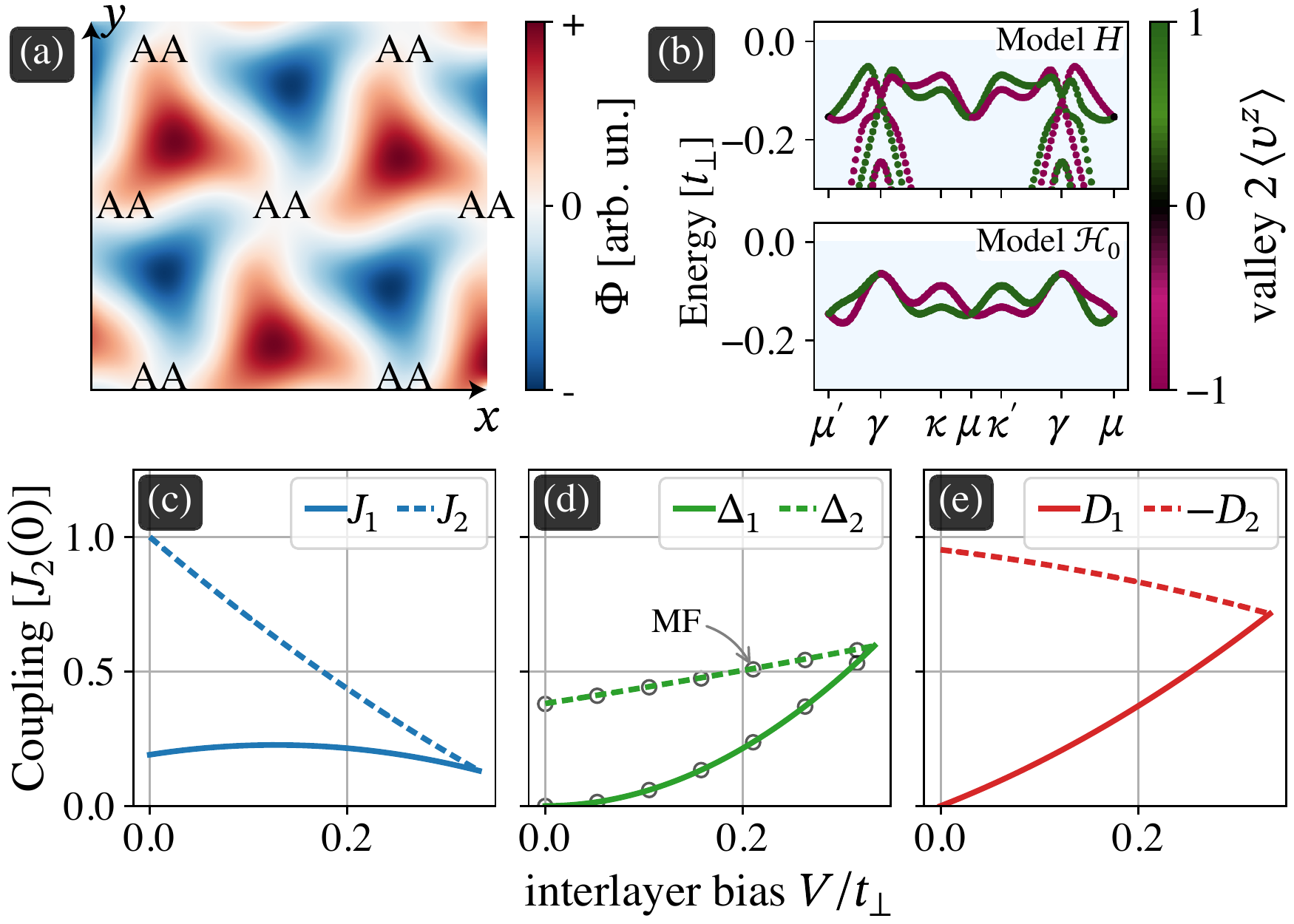}
\caption{\label{fig:fig3}
Effect of interlayer bias $V$ on single-particle properties and effective
valley--valley exchange interactions in the anisotropic valley-Heisenberg
model~$\mathcal{H}_{\vop}$~\eqref{eq:Heisenberg}.
(a)~Local valley (Berry) flux $\Phi(\vec{r},E)$ near half-filling (at energy
$E\simeq -0.1 t_\perp$), averaged over the microscopic scale of
model~\eqref{eq:fullModel} including both layers, see Eq.~\ref{eq:valleyflux}.
The staggered flux is largest in AB/BA regions and vanishes in the AA regions.
(b)~Flat band as obtained from the microscopic model~\eqref{eq:fullModel} (top
panel) compared with the phenomenological model~\eqref{eq:effmodel} (bottom
panel) at finite interlayer bias $V=0.33t_\perp$, and with
${\gamma_1=\gamma_2=0.07 t_\perp}$ and ${\phi_1=-\phi_2=0.7}$. Note the
difference with the $V=0$ result in Fig.~\ref{fig:fig2}(c).
(c--e) Isotropic ($J_n$), anisotropic ($\Delta_n$) and antisymmetric ($D_n$)
valley exchange-couplings, cf.~Eq.~\eqref{eq:Heisenberg}, for first and second
neighbors ($n=1,2$) as interlayer bias $V$ increases. The interlayer bias can
enhance the first-neighbor coupling even to the point where it has the same
magnitude and phase as the second-neighbor coupling (here $V\simeq 0.3
t_\perp$). Panel~(d) shows the numerical mean field result (open circles)
superimposed on top of the analytical result (solid/dashed lines)
\cite{supmat}.
}
\end{figure}
%%%%%%%%%%%%%%%%%%%%%%%%%%%%%%%%%%%%%%%%%%%%%%%%%%%%%%%%%%%%%%%%%%%%%

Including a finite interlayer bias ${V>0}$ in Eq.~\eqref{eq:fullModel} induces
effective valley-dependent fluxes $\val \Phi(\vec{r}_i,E)$ in real space that
remove the valley degeneracy, see Fig.~\ref{fig:fig3}(b); within the
low-energy model $\mathcal{H}_0$ \eqref{eq:effmodel}, they modify the
valley--orbit phases $\phi_1$ and $\phi_2$.  This is formalized by defining
the valley flux of low-energy states~\cite{ChenLee2011PRB,
WolfZilberberg2019PRL, ManescoRodrigues2020ae} near the energy $E$ and at
position~$\vec{r}_i$ as
\begin{align} \label{eq:valleyflux}
\Phi(\vec{r}_i,E) = \int_{\textrm{BZ}} \frac{d^2 \vec k}{(2\pi)^2}
\frac{\epsilon_{\alpha \beta}}{2} \langle \vec r_i  | G\,
(\partial_{k_\alpha}G^{-1}) \,(\partial_{k_\beta}G)| \vec r_i \rangle,
\end{align}
where $G = [E - H(\vec k)+i0^+]^{-1} \mathcal{P}$ is the valley Green's
function with valley-polarization operator ${\mathcal{P}=2v^z}$, and
$\epsilon_{\alpha\beta}$ denotes the Levi-Civita symbol.  For our flat
bands, we find that the interlayer bias induces a staggered valley flux, see
Fig.~\ref{fig:fig3}(a). This flux can be included in the low-energy model
$\mathcal{H}_0$~\eqref{eq:effmodel}, through a Peierls substitution, i.e.,
$\gamma_n\mapsto\gamma_n(V)\,e^{i \sigma^z \phi_{n}(V)}$ for $n=1,2$, cf.
Fig.~\ref{fig:fig2}(b). It contributes dominantly to $\phi_1$, and provides an
additional correction to $\phi_2$ accounting for the tilt in the pattern.  The
bands of the effective model $\mathcal{H}_0(V)$ obtained in this way
qualitatively agree with the bands of the atomistic tight-binding
Hamiltonian~\eqref{eq:fullModel} evaluated at a finite interlayer bias $V$, see
Fig.~\ref{fig:fig3}(b).

%% Note: There is an alternative version for this paragraph commented out 
%%       below 
Consequently, the interlayer bias directly controls the effective
valley-exchange couplings in model $\mathcal{H}_v$~\eqref{eq:Heisenberg}
through the induced valley--orbit couplings $\phi_{1}(V)$ and $\phi_{2}(V)$. In
Figs.~\ref{fig:fig3}(c-f), we see that the couplings $J_1$, $\Delta_2$, and
$D_2$ do not change significantly with increasing bias $V$, while the coupling
$J_2$ decreases substantially, and $\Delta_1$ and $D_1$ both turn finite and
increase appreciably. As a result, we find here that the interlayer bias (i)
increases the easy-plane exchange anisotropy (increasing $\Delta_n$), (ii)
decreases the overall tendency for anti-valleymagnetic order and geometric
frustration (decreasing $J_n$), and (iii) increases canting (through $D_n$).
Interestingly, this means that the interlayer bias switches between the two
mechanisms responsible for valley spirals. Note that there is also a
competition of canting between first-neighbor and second-neighbor orbital pairs
that influences the length scale of the valley spiral, where in numerical
mean-field calculations we predominantly observed $120^{\circ}$ and
$60^{\circ}$ spiral structures. A more detailed analysis of competing spiral
structures is beyond the scope of this work.

% % % % % % % % % % % % % % % % % % % % % % % % % % % % % % % % % %
% % % % % % % % Signatures and limitations

%%%%%%%%%%%%%%%%%%%%%%%%%%%%%%%%%%%%%%%%%%%%%%%%%%%%%%%%%%%%%%%%%%%%%
\begin{figure}%[t!]%[!h]
\includegraphics[width=\columnwidth]{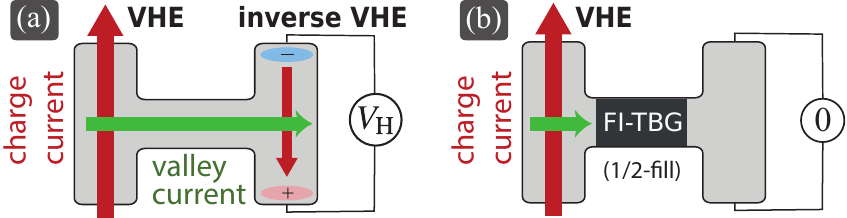}
\caption{\label{fig:fig4}
Schematic setup for an experiment signaling the presence of a valley
spiral state.
(a) Standard four-terminal device, with valley Hall effect (VHE) driven by a
charge current and inverse VHE driven by a valley current and producing a
finite voltage~$V_{\textrm{H}}$~\cite{XiaoNiu2007PRL, YamamotoTarucha2015JPS}.
(b) FI-encapsulated TBG (FI-TBG) at half-filling of the flat band acts as
filter blocking the valley current and suppresses the voltage~$V_{\textrm{H}}$
of the inverse VHE.
}
\end{figure}
%%%%%%%%%%%%%%%%%%%%%%%%%%%%%%%%%%%%%%%%%%%%%%%%%%%%%%%%%%%%%%%%%%%%%

In contrast to spin, valley is an orbital degree of freedom, and thus provides
an extra challenge when it comes to interpretation and experimental
verification of valley-physics~\cite{XiaoNiu2007PRL,JiangGuinea2013PRL,
LinnikLinnik2014PRB, YamamotoTarucha2015JPS,
MartinyJauho2019PRB,NguyenCharlier2016PRL, AkhmerovBeenakker2007PRL,
SanJosePrada2013PRB,RickhausEnsslin2018NL, XuGeim2019NC, HuangLeRoy2018PRL}. A
promising direction is to make use of the valley Hall effect (VHE), where band
electrons from valley $K$ flow in the opposite direction as those from valley
$K'$, leading to transverse charge-neutral valley
currents~\cite{XiaoNiu2007PRL, ShimazakiTarucha2015NP, YamamotoTarucha2015JPS}.
These currents can be detected as they induce voltages in other
regions of the material through the inverse-VHE, see Fig.~\ref{fig:fig4}(a).
Such a four-terminal transport setup enables the detection of our
valley-correlated state, i.e., the latter can be characterized through its
action on a valley Hall measurement when embedding our system into a suitable
device geometry, see Fig.~\ref{fig:fig4}(b). For example, a valley-magnet
($\langle\vop_I^z\rangle\neq 0$) acts as a valley-filter and can be used to
suppress the valley Hall signal for one valley but not the other. In our case,
we expect the planar valley-spiral ($\langle\vop_I^z\rangle=0$) to act as a
``coherent valley mixer''~\cite{MengHe2012PRB, YanWu2016PRB,
MorpurgoGuinea2006PRL}. This would strongly suppress the valley Hall signal
when the chemical potential is swept to approach half-filling of the flat band,
thus providing an experimental signature by which to detect the valley spiral.

To conclude, our results put forward a minimal graphene-based heterostructure
displaying spontaneous valley-mixing, opening up a pathway to explore
valley-correlated states in twisted graphene multilayers. Going beyond this
work, FI-encapsulated TBG and twisted \emph{double}-bilayer graphene (TDBG)
have analogous electronic band structures, except that spin in the former
replaces the additional graphene layer in the latter. This can be understood by
considering monolayer graphene on a magnetic substrate compared with isolated
bilayer graphene. This similarity suggests that many recent proposals and
observations~\cite{ShenZhang2020NP, LiuKim2019ae, HeYankowitz2020ae} for the
latter may also apply to the model studied here. In particular, besides
correlated insulating states \cite{BurgTutuc2019PRL, CaoJarilloHerrero2020N},
ferromagnetic superconductors emerge in TDBG~\cite{LiuKim2019ae}, which, by
extension, could lead to valleymagnetic superconductivity in our model
when doped away from half-filling. Ultimately, the proposed FI-TBG can become a
potential candidate to realize valley-analogous versions of fractional quantum
Hall states~\cite{AbouelkomsanBergholtz2020PRL, LiuBergholtz2020ae,
LedwithVishwanath2020PRR, RepellinSenthil2020PRR}, and quantum valley-liquids
in twisted van der Waals materials~\cite{WuXu2019PRB, IrkhinSkryabin2018JL,
GonzalezArragaSanJose2017PRL, NatoriAndrade2019PRB}.

\begin{acknowledgments}
We acknowledge financial support from the Swiss National Science
Foundation. J.L.L. acknowledges the computational resources provided
by the Aalto Science-IT project.
\end{acknowledgments}

\bibliography{paper.bib}

%% file: main_supmat.tex
%!TeX root = ms.tex

\section{Microscopic tight-Binding model}

For the reader's convenience, here we repeat the generic atomistic
tight-binding Hamiltonian for stacked graphene [see Eq.~(1) in the main text]
with atoms located at coordinates $\lbrace \vec{r}_i \rbrace$, i.e.,
\begin{align}
    H_0 = \sum_{i\neq j,s} t(\vec{r}_i-\vec{r}_j) \, c^\dagger_{i s} c^\pdagger_{j s} + \sum_{i s} V(\vec{r}_i) \, c^\dagger_{i s} c^\pdagger_{i s},
\end{align}
where $c^{(\dagger)}_{is}$ destroys (creates) an electron at site $\vec{r}_i$
with spin $s\in\lbrace\pm 1/2\rbrace$. The hopping amplitudes can be
parametrized as Slater-Koster transfer integrals between the atomic
orbitals~\cite{SM_MoonKoshino2013PRB,SM_WolfZilberberg2019PRL}, i.e.,
\begin{align}
 - t (\vec{R}) = t_{p p \pi}(R) \cdot \left( 1 - \left( \frac{\vec{R}
   \cdot \hat{z}}{R} \right)^2 \right) + t_{p p \sigma}(R) \cdot \left(
   \frac{\vec{R} \cdot \hat{z}}{R} \right)
\end{align}
with decaying overlap amplitudes $t_{p p \pi} = t\, e^{- (R - a_0) / \ell}$ and
$t_{p p \sigma} = t_\perp \, e^{- (R - d) / \ell}$ where $a_0 = a / \sqrt{3}
\approx 0.142$ nm is the intralayer interatom distance, $d \approx 2.35 a_0$ is
the interlayer spacing, $t \approx 2.7$ eV is the first-neighbor transfer
integral and $t_\perp \approx - 0.18 t$ is interlayer transfer
integral, and $\ell \approx 0.33 a_0$ is the decay length of the overlap
integrals. In our case, the onsite potential
\begin{align}
 V(\vec{r}) = \mu + \mathrm{sign}(z_i) \, V
\end{align}
contains the overall chemical potential $\mu$ and the interlayer bias $V$.

As explained in the main text, the presence of a ferromagnetic insulator (FI)
introduces an effective exchange field, such that the the electron spin couples
to an effective magnetic moment $\vec{m}(\vec{r})$, i.e.,
\begin{align}
    H_{\labelJ} = \sum_{j ss'} (\vec{m}(\vec{r}_j)\cdot\vec{\sigma})_{ss'} c^\dagger_{j s} c_{j s'}^\pdagger,
\end{align}
where $\vec{\sigma}=(\sigma_x,\sigma_y,\sigma_z)$ are the spin Pauli matrices.
We assume that the FI is layer-antiferromagnetic, i.e.,
$\vec{m}(\vec{r})=\mathrm{sign}(z_i) m\,\hat{z}$. The FI also induces the
Rasbha spin--orbit interaction
\begin{align}
    H_{\labelR} = \sum_{\langle i j \rangle, s s'} i \lambdaR(\vec{r}_i) (\vec{\sigma}\times\vec{d}_{ij})_{s s'}^z \, c^\dagger_{i s} c^\pdagger_{j s'},
\end{align}
where $\vec{d}_{ij}$ is the bond vector connecting intralayer sites $i,j$. The
Rasbha spin--orbit coupling in this case is
$\lambdaR(\vec{r})=\mathrm{sign}(z_i) \lambdaR$.

\begin{figure}
\centering
\includegraphics[width=3.8in]{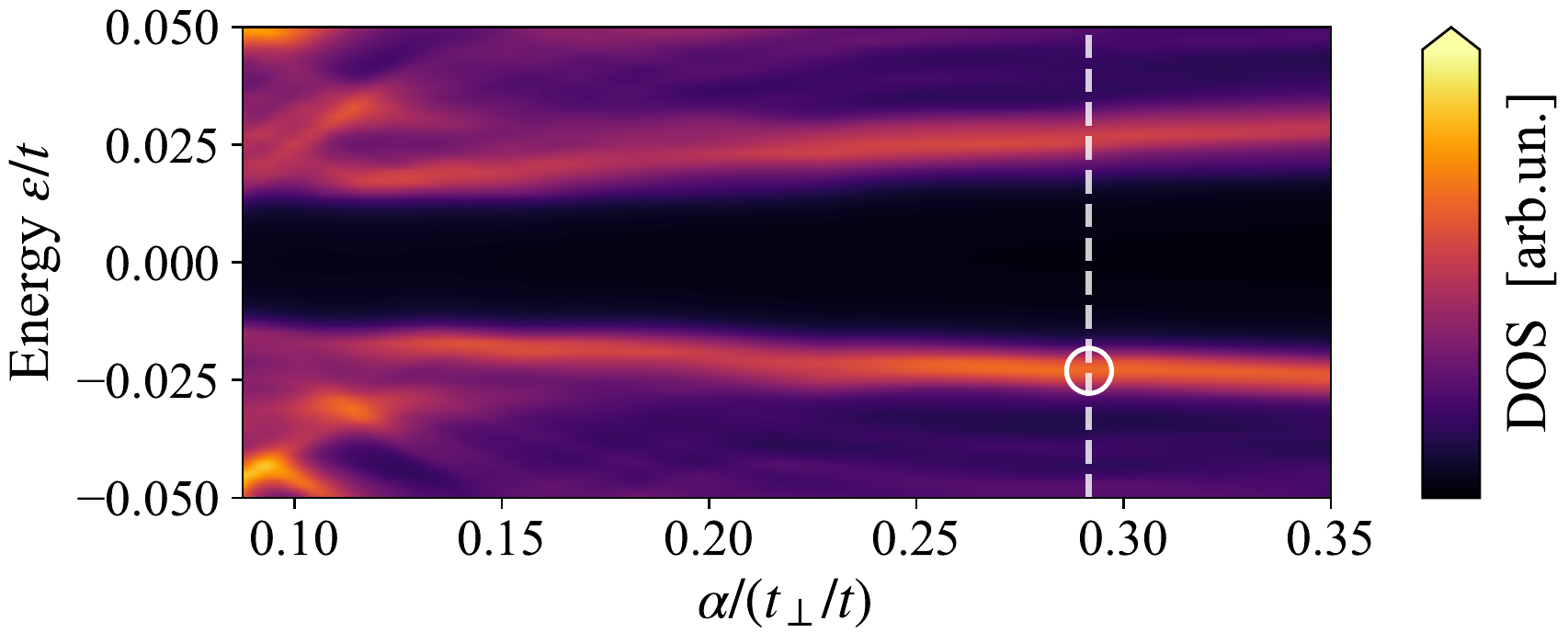}
\caption{\label{sfig1}
Density of states of FI-encapsulated twisted bilayer graphene for different
interlayer hopping amplitudes as a function of $\alpha/t_\perp$ (calculated
with $\alpha\approx 2^{\circ}$ fixed, $t_\perp/t\in[0.1,0.4]$) with
$m=\lambdaR=0.33 t_\perp$. The dashed line and circle indicate the van Hove
singularity associated to the flat band investigated in the main text.
}
\end{figure}

If we now consider the Hamiltonian $H=H_0+H_{\labelJ}+H_{\labelR}$ twisted
bilayer graphene at fixed physical parameters $t$, $t_\perp$, $m$ and
$\lambdaR$, the electronic spectrum still depends on the twist angle
$\alpha$. We can investigate this dependence by considering the density of
states, to identify van Hove singularities and band gaps, see Fig.~\ref{sfig1}.
Note that it is generally computationally expensive to vary the twist angle
in the tight-binding calculation. However, a rescaling argument in the
parameter $\alpha/t_\perp$ can be used to vary the interlayer hop amplitude at
fixed angle instead~\cite{SM_WolfZilberberg2019PRL}.

%%%%%%%%%%%%%%%%%%%%%%%%%%%%%%%%%%%%%%%%%%%%%%%%%%%%%%%%%%%%%%%%%%%%%%%%%%%%%%%
%%%%%%%%%%%%%%%%%%%%%%%%%%%%%%%%%%%%%%%%%%%%%%%%%%%%%%%%%%%%%%%%%%%%%%%%%%%%%%%

%%%%%%%%%%%%%%%%%%%%%%%%%%%%%%%%%%%%%%%%%%%%%%%%%%%%%%%%%%%%%%%%%%%%%%%%%%%%%%%
%%%%%%%%%%%%%%%%%%%%%%%%%%%%%%%%%%%%%%%%%%%%%%%%%%%%%%%%%%%%%%%%%%%%%%%%%%%%%%%
%%%%%%%%%%%%%%%%%%%%%%%%%%%%%%%%%%%%%%%%%%%%%%%%%%%%%%%%%%%%%%%%%%%%%%%%%%%%%%%
\section{Fermi-Hubbard model}

\subsection{Hamiltonian}
In the main text, we consider a generalized Fermi-Hubbard model describing
hopping between effective electronic orbitals that are punished by local
on-site repulsion~[c.f. Eqs. (2) and (3) in the main text]. In our case, the
spin degree of freedom is hybridized and valleys $K,K'$ take the role of a
pseudospin degree of freedom, which we will denote with $v=\pm1/2$. The
corresponding Hamiltonian is
\begin{align} \label{eq:HubbardSupmat}
   \mathcal{H} = \mathcal{H}_t + \mathcal{H}_U = \sum_{i \neq j, \val=\pm 1/2} t^{\val}_{i j} \; d_{i
   \val}^{\dagger} d_{j \val}^\pdagger + \frac{U}{2} \sum_{i,v} n_{iv} n_{i(-v)},
\end{align}
where $n_{i \val} = d^{\dagger}_{i \val} d_{i \val}^\pdagger$ is the
local number operator with creation/annihilation operators $\{ d_{i
\val}^\pdagger, d^{\dagger}_{j \val'} \} = \delta_{i j} \delta_{\val
\val'}$, and $U > 0$ is the Hubbard interaction strength. In particular, we
allow valley-dependent hopping amplitudes $t_{i j}^{\val} = t_{i
j}^{\mathrm{s}} + \val \, t_{i j}^{\mathrm{a}}$.

% % % % % % % % % % % % % % % % % % % % % % % % % % % % % % % % % % % % % % % %
\subsection{Symmetry}

The configuration of the ferromagnetic insulators in the microscopic model (see
main text) allows us to introduce the combination of time reversal and
structural symmetry operations as pseudo-time-reversal symmetry operation. In
our Fermi-Hubbard model, this symmetry operation is then given by $\mathcal{T}
= i \sigma_y  \mathcal{K}$, where $\mathcal{K}$ denotes complex conjugation.
This symmetry implies $t_{i j}^{\val} = (t_{i j}^{-\val})^*$, which is
equivalent to
\begin{align}
   t_{i j}^{\val} = \gamma_{i j} e^{i \val \phi_{i j}} \qquad \text{with}
   \qquad \gamma_{i j} = \gamma_{j i} > 0, \quad \phi_{i j} = - \phi_{j i} \in
   [0, 2 \pi]
\end{align}
due to hermaticity.
These hopping amplitudes lend themselves to an interpretation as
pseudospin--orbit coupling (or ``valley--orbit'' coupling), which can be seen
for example by looking at the Kane-Mele model~\cite{SM_KaneMele2005PRL}. In the
main text, we restricted ourselves to first- and second-neighbor amplitudes,
i.e.,
\begin{align*}
t_{\langle i j \rangle}^{\val} = \gamma_1 \, e^{i \val \varepsilon_{i j} \phi_1} \quad \text{and} \quad t_{\langle \langle i j \rangle \rangle}^{\val} = \gamma_2 \, e^{i \val \nu_{i j} \phi_2},\qquad \text{where} \qquad \nu_{i j}\in\lbrace\pm 1, 0\rbrace
\end{align*}
is antisymmetric and also restricted by structural symmetries.

Introducing the spinor $\psi_i = (d_{i \uparrow}, d_{i \downarrow})$, we can
also define spin operators $v_{i \alpha} = (1 / 2) \psi^{\dagger}_i
\sigma_{\alpha} \psi_i$, where $\sigma_{\alpha}$ are the Pauli matrices \
($\alpha = x, y, z$). These transform as $\vec{v}_i \mapsto R \vec{v}_i$ under
spinor rotations $\psi_i \mapsto U \psi_i = \exp (i \varphi \vec{n} \cdot
\vec{\sigma} / 2) \psi_i$, where $R$ is the spin rotation associated to the
spinor rotation $U$. Our Hamiltonian $H$ then has the symmetry axis $\vec{n} =
\vec{e}_z$. Furthermore, the mirror operation $U = i \sigma_x$ is a symmetry if
$t_{i j}^{\val} = t_{i j}^{-\val}$. The latter paired with time reversal
symmetry implies $\phi_{i j} = 0, \pm \pi / 2, \pi$. Note that this mirror
operation is generally \textit{not} a symmetry of our Hamiltonian.

% % % % % % % % % % % % % % % % % % % % % % % % % % % % % % % % % % % % % % % %
\subsection{Hartree-Fock mean field approximation}
We introduce the mean density matrix $\rho_{\val \val'}^i = \langle
d^{\dagger}_{i \val} d_{i \val'}^\pdagger \rangle$ and use the mean field
approximation~\cite{snote1}
to find
\begin{align} 
   \mathcal{H}_U \approx \mathcal{H}_U (\rho) 
 &= U \sum_{i \val} \underbrace{\rho_{(-\val)
   (-\val)}^i c_{i \val}^{\dagger} c_{i \val}^\pdagger}_{\text{Hartree}} -
   \underbrace{\rho_{(-\val)\val}^i c_{i \val}^{\dagger} c_{i
   (-\val)}^\pdagger}_{\text{Fock}} + \underbrace{\frac{U}{2} \sum_{i v}
   \rho_{\val\val}^i \rho_{(-\val)(-\val)}^i - | \rho_{\val(-\val)}^i |^2}_{\equiv \sum_i E_0 (\rho^i)} \nonumber \\
 &= \sum_{i\val\val'} \bar{U}_{\val\val'}(\rho^i)\; c_{i \val}^{\dagger} c_{i \val'}^\pdagger + \sum_i E_0 (\rho^i),
\end{align}
where $\bar{U}_{\val\val'}(\rho^i)=U\,\delta_{\val\val'}
\rho_{(-\val) (-\val)}^i - U\, \delta_{\val(-\val')}
\rho_{(-\val)\val}^i $. The density matrix $\rho^i$ is then obtained
through the self-consistency relation
\begin{align*}
\rho_{\val \val'}^i = \langle
d^{\dagger}_{i \val} d_{i \val'}^\pdagger \rangle \approx Z^{- 1} (\rho) \; \mathrm{tr}
\left[e^{- \beta \mathcal{H}(\rho)} c^\dagger_{i \val} c^\pdagger_{i \val'}\right],
\end{align*}
which must be solved numerically (e.g., through fixed point iteration). The
expectation value of the valley operator can then be extracted by observing
that
$
	\rho^i = \frac{1}{2} \left( \langle n_i \rangle + \langle \vec{v}_i \rangle \cdot \vec{\sigma} \right),
$
where $n_i = \sum_v n_{iv}$ is the occupation number at site $i$.

%%%%%%%%%%%%%%%%%%%%%%%%%%%%%%%%%%%%%%%%%%%%%%%%%%%%%%%%%%%%%%%%%%%%%%%%%%%%%%%
%%%%%%%%%%%%%%%%%%%%%%%%%%%%%%%%%%%%%%%%%%%%%%%%%%%%%%%%%%%%%%%%%%%%%%%%%%%%%%%
%%%%%%%%%%%%%%%%%%%%%%%%%%%%%%%%%%%%%%%%%%%%%%%%%%%%%%%%%%%%%%%%%%%%%%%%%%%%%%%
\section{Effective valley--valley exchange interactions}

In the large-$U$ limit of the Fermi-Hubbard model $\mathcal{H}$, the hoppings
$\mathcal{H}_t$ can be included in second-order perturbation theory, or
equivalently by using the Schrieffer-Wolff transformation that eliminates the
hoppings to first order via the canonical transformation
\cite{SM_AltlandSimons2010book,SM_Spalek2007arxiv}
\[ \mathcal{H}' = e^{- A}  \mathcal{H} e^A \approx \mathcal{H} - [A, \mathcal{H}] + \frac{1}{2}  [A, [A, \mathcal{H}]] + \ldots \approx \mathcal{H}_U - \frac{1}{2} \left[A,\mathcal{H}_t\right] + \mathcal{O}(t^3),
\]
where $A$ is chosen such that $\mathcal{H}_t - \left[ A, \mathcal{H}_U \right]
= 0$. The last constraint is solved in the subspace of states for which each
lattice site is singly occupied (most relevant for large $U$ at half-filling).
Denoting the corresponding subspace projector $P$ (and its orthogonal
complement $P_\perp$), one can show that $A=(P \mathcal{H}_t P_\perp - P_\perp
\mathcal{H}_t P)/U$ leads to
\begin{align}
	\mathcal{H}_v \equiv P H' P = - P H_t^2 P = - P \sum_{i \neq j, \val \val'} \frac{t^{\val}_{i j}
(t^{\val'}_{i j})^*}{U} d_{j
\val'}^{\dagger} d_{i \val'}^\pdagger d_{i \val}^{\dagger} d_{j \val}^\pdagger P,
\end{align}
which after some manipulations and using $v_{i+}=
d^\dagger_{i(1/2)}d_{i(-1/2)}^\pdagger=v_{ix}+i v_{iy}=(v_{i-})^\dagger$ and
$v_{iz}=(n_{i(1/2)}-n_{i(-1/2)})/2$ leads to an anisotropic Heisenberg model
with antisymmetric exchange, i.e.,
\[ H_v = \sum_{i \neq j, \val} J_{i j}  \vec{v}_i \cdot
   \vec{v}_j + \Delta_{i j} v_{i z} v_{j z} + D_{i j}  \vec{z} \cdot
   (\vec{v}_i \times \vec{v}_j) + \text{const} ., 
\]
with the exchange couplings
%S
\begin{align}\label{eq:couplings}
	J_{i j} &= J_{i j}^0 \pm \frac{\Delta_{i j}}{2} = \frac{2 \left| t_{i j}^{\mathrm{s}} \right|^2}{U} - \frac{2 \left| t_{i j}^{\mathrm{a}} \right|^2}{U} = \frac{2 \gamma_{i j}^2}{U}  (\cos^2 \phi_{i j} - \sin^2 \phi_{i j}), \\
	\Delta_{i j} &= \frac{4 \left| t_{i j}^{\mathrm{a}}
   \right|^2}{U} = \frac{4 \gamma_{i j}^2}{U} \sin^2 \phi_{i j}, \\
   D_{i j} &= \frac{4 \Im \left[ t_{i j}^{\mathrm{s}} \left( t_{i j}^{\mathrm{a}} \right)^* \right]}{U} = \frac{2 \gamma_{i j}^2}{U} \sin (2 \phi_{i j})
   . 
\end{align}
Note that this Hamiltonian $\mathcal{H}_v$ is compatible with the spinor
rotation symmetry $U_{\varphi} = \exp (i \varphi \sigma_z / 2)$, i.e., around
the axis $\vec{n} = \vec{e}_z$ -- just like the full Hamiltonian
$\mathcal{H}$~\eqref{eq:HubbardSupmat}. Additionally imposing mirror symmetry
in spinor space would lead to $\phi_{i j} = 0, \pi$ with $J_{i j} = 2 \gamma_{i
j}^2 / U$ and $\Delta_{i j} = D_{i j} = 0$ or else $\phi_{i j} = \pm \pi / 2$
with $J_{i j} = - 2 \gamma_{i j}^2 / U$, $\Delta_{i j} = 4 \gamma_{i j}^2 / U$
and $D_{i j} = 0$.

% % % % % % % % % % % % % % % % % % % % % % % % % % % % % % % % % % % % % % % %
\subsection{Extracting effective valley exchange couplings from mean field}

The effective exchange couplings $J_{i j}$, $\Delta_{i j}$ and $D_{i j}$
derived in the strong-$U$ limit, see Eq.~\eqref{eq:couplings}, can also be
extracted directly from the Hubbard model~\eqref{eq:HubbardSupmat} using
numerical mean field calculations. To this end, we compare ground state
energies of the same trial states (i.e. valley-polarized in-plane,
valley-polarized out-plane and spin spirals). For example, the difference
between ground state energies of trial states polarized in-plane and those
out-plane yields the anisotropic coupling $\Delta_{i j}$. The other couplings
can be obtained in a similar fashion.

%% file: ms.bbl
\begin{thebibliography}{83}
\bibitem [{not()}]{note1}%
  \BibitemOpen
  \bibinfo {note} {We introduce the nomenclature (anti)valleymagnetic for the
  valley sector in distinction to (anti)ferromagnetism in the spin
  sector.}\BibitemShut {Stop}%
\end{thebibliography}
\begin{thebibliography}{6}
\bibitem [{sno()}]{snote1}%
  \BibitemOpen
  \bibinfo {note} {Note that we use $c_{\mu}^{\dagger} c_{\nu}^{\dagger}
  c_{\nu} c_{\mu} \approx \rho_{\mu \mu} c_{\nu}^{\dagger} c_{\nu} + \rho_{\nu
  \nu} c_{\mu}^{\dagger} c_{\mu} - \rho_{\mu \nu} c_{\nu}^{\dagger} c_{\mu} -
  \rho_{\nu \mu} c_{\mu}^{\dagger} c_{\nu} - \rho_{\mu \mu} \rho_{\nu \nu} +
  \rho_{\mu \nu} \rho_{\nu \mu}$, where $\rho_{\mu \nu} = \langle
  c^{\dagger}_{\mu} c_{\nu} \rangle$ for indices $\mu$, $\nu$.}\BibitemShut
  {Stop}%
\end{thebibliography}
